\documentclass{iopart}

\usepackage{amsfonts}
\usepackage{amssymb}
\usepackage[english]{babel}
\usepackage{exscale}
\usepackage[final]{graphicx}
\usepackage{mathpple}

\newcommand{\ignore}[1]{}

\newcommand{\href}[2]{#2}

\begin{document}

\title{Finding apparent horizons and other two-surfaces of constant expansion}
\author{Erik Schnetter}
\address{
   Institut f\"ur Astronomie und Astrophysik, Auf der Morgenstelle,
   Universit\"at T\"ubingen, D-72076 T\"ubingen, Germany}
\ead{schnetter@uni-tuebingen.de}

\begin{abstract}
   Apparent horizons are structures of spacelike hypersurfaces that
   can be determined locally in time.  Closed surfaces of constant
   expansion (CE surfaces) are a generalisation of apparent horizons.
   I present an efficient method for locating CE surfaces.  This
   method uses an explicit representation of the surface, allowing for
   arbitrary resolutions and, in principle, shapes.  The CE surface
   equation is then solved as a nonlinear elliptic equation.

   It is reasonable to assume that CE surfaces foliate a spacelike
   hypersurface outside of some interior region, thus defining an
   invariant (but still slicing-dependent) radial coordinate.  This
   can be used to determine gauge modes and to compare time evolutions
   with different gauge conditions.  CE surfaces also provide an
   efficient way to find new apparent horizons as they appear e.g.\ in
   binary black hole simulations.
\end{abstract}


\submitto{\CQG}

\section{Introduction}

   Apparent horizons serve several purposes during the numerical
   evolution of general relativistic spacetimes.  In vacuum, spacetime
   has no other prominent features that can be determined locally in
   time, such as e.g.\ the shock waves found in hydrodynamics.  Event
   horizons are global structures, and it is not possible to find them
   during a time evolution, as the future of the spacetime is not yet
   known.  Yet an apparent horizon indicates, under reasonable
   assumptions, that there is an event horizon present at or outside
   the apparent horizon.  Furthermore, apparent horizons are also
   under certain circumstances isolated horizons, making it then
   possible to calculate their mass and spin \cite{generic-ih,
   rotating-ih, dreyer2003}.

   Locating apparent horizons is also necessary when one wants to
   apply excision boundary conditions in a numerical time evolution,
   where the apparent horizon is used to determine the location and
   shape of the excised regions.  It is therefore important to have a
   fast and robust apparent horizon finder that is closely integrated
   with the evolution code.

   Apparent horizons can be generalised to closed two-surfaces of
   constant expansion (CE surfaces).  Empirically, CE surfaces lead to
   a foliation of a spacelike hypersurface outside of some interior
   region.  Often, the gauge in which numerical solutions to
   Einstein's equations are obtained is itself the result of time
   evolution equations, so that the gauge condition is not known
   explicitly.  CE surfaces can be used to define a radial coordinate
   that is independent of the spatial gauge choice.  This allows to
   determine gauge modes and compare different gauge conditions.

   This article is structured as follows.
   Section~\ref{sec:definitions} introduces the notation and gives the
   defining equations for apparent horizons and surfaces of constant
   expansion.  Section~\ref{sec:numerics} explains the discretisation
   of the surfaces and the numerical methods used to solve the
   equations.  Section~\ref{sec:tests} shows tests with analytic
   solutions and compares the finder's accuracy and speed to existing
   implementations of apparent horizon finders.  Finally,
   section~\ref{sec:applications} presents hypersurface foliations and
   apparent horizon pre-tracking as applications for CE surfaces, and
   section~\ref{sec:summary} summarises the results.

\section{Definitions}
\label{sec:definitions}

   A marginally outer-trapped surface is a two-surface within a
   spacelike hypersurface with zero expansion and spherical topology.
   An apparent horizon is the outermost such surface.  The condition
   that it be an outermost such surface is difficult to verify in
   practice, and I will follow the current usage to disregard this
   condition.

   Therefore, at each point of an apparent horizon, the condition
   $H=0$ has to hold, where
\begin{equation}
\label{eq:ah-func}
   H := \nabla_i s^i + K_{ij} (s^i s^j - \gamma^{ij})
\end{equation}
   is the so-called \emph{apparent horizon function} (see e.g.\
   \cite{ah-finding, alcubierre2000}).  Here $s^i$ is the (spacelike)
   outward normal to the horizon, and $\gamma_{ij}$ and $K_{ij}$ are
   the three-metric and extrinsic curvature of the spacelike
   hypersurface.

\begin{figure}
\begin{center}
{\footnotesize(a)}\includegraphics[scale=1.1]{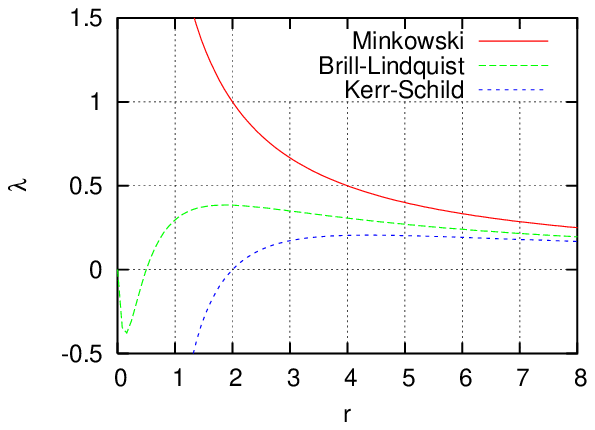}
\hspace{2em}
{\footnotesize(b)}\includegraphics[scale=1.1]{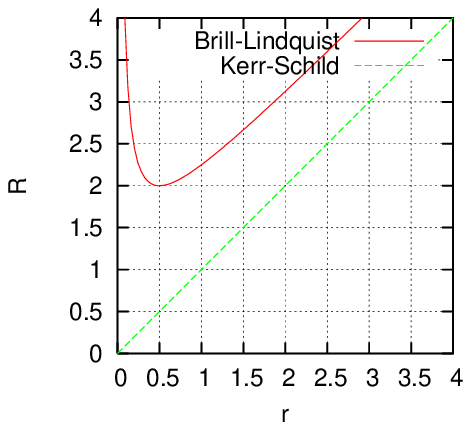}
\end{center}
\caption{
   (a) Expansion $\lambda$ vs.\ coordinate radius $r$ for the
   Minkowski, a Brill--Lindquist ($\mu=1$), and a Kerr--Schild ($M=1$)
   metric.  All curves tend to $\lambda = 2/r$ for $r \to \infty$.
   Apparent horizons are located where $\lambda=0$.
   (b) Areal radius $R$ vs.\ coordinate radius $r$ for a
   Brill--Lindquist ($\mu=1$) and a Kerr--Schild ($M=1$) metric.  Note
   that for Kerr--Schild coordinates $R=r$.  The apparent horizons are
   here located where $R=2$.}
\label{fig:lambda}
\label{fig:radius}
\end{figure}

   A generalisation of apparent horizons, which have zero expansion,
   are surfaces that have instead a certain constant expansion
   $\lambda$.  These CE surfaces are defined through a condition
   $H_\lambda=0$, where
\begin{equation}
\label{eq:ce-func}
   H_\lambda := H - \lambda
\end{equation}
   is the \emph{CE surface function}.  Apparent horizons are CE
   surfaces with $\lambda=0$.  In asymptotically spacelike foliations
   of asymptotically flat spacetimes, coordinate spheres with $r \to
   \infty$ are CE surfaces with $\lambda = 2/r$.
   Figure~\ref{fig:lambda}~(a) shows as examples the behaviour of
   $\lambda$ vs.\ $r$ for coordinate spheres in some spherically
   symmetric spacetimes.

   In the case of a vanishing extrinsic curvature, CE surfaces are
   also constant mean curvature surfaces (CMC surfaces).  These lead
   under certain assumptions, such as a positive energy density, to a
   foliation of the spacelike hypersurface outside a small interior
   region \cite{huisken1996}.  It is not unreasonable to assume the CE
   surfaces will also lead to a foliation under similar conditions,
   although this has not yet been proven.

   It is clear from figure~\ref{fig:lambda}~(a) that the expansion
   $\lambda$ is not monotonic and therefore cannot be used to define a
   radial coordinate.  However, such an invariant radial coordinate
   would be most useful for numerical simulations, e.g.\ to compare
   results obtained with different gauge conditions, or to have some
   insight into the coordinate distortions caused by a gauge
   condition.

   If the CE surfaces have monotonically increasing areas, then it is
   natural to consider the areal radius
\begin{equation}
\label{eq:areal-radius}
   R = \sqrt{A / 4 \pi}
\end{equation}
   of a surface, where $A$ is the area of the surface as measured
   using the three-metric in the spacelike hypersurface.  This areal
   radius is twice the irreducible mass for an apparent horizon.  (It
   is also equal to the radial coordinate $r$ when the radial gauge
   $\gamma_{\theta\theta} = r$ is used in spherical symmetry.)
   Figure~\ref{fig:radius}~(b) shows as examples the behaviour of the
   areal radius of coordinate spheres for some spherically symmetric
   spacetimes.

   In order for $R$ to be a useful coordinate in practice, it is
   necessary that one be able to find a CE surface with a specific
   areal radius $R$.  It is not always possible to locate such a
   surface by considering CE surfaces with varying values of
   $\lambda$, because varying $\lambda$ is ill-defined where the
   gradient of $\lambda$ vanishes (see again
   figure~\ref{fig:lambda}~(a)).  Instead, I define the
   \emph{areal-radius-corrected CE surface function}
\begin{equation}
\label{eq:ar-func}
   H_r := H - \bar H + (R - r) / r
\end{equation}
   which is zero if and only if the trial surface is a CE surface with
   the areal radius $r$.  Here $\bar H$ is the average of the
   expansion of the surface, i.e.\ $\bar H = \int d\Omega\, H(\Omega)
   / 4 \pi$.  $R$ is the surface's areal radius as defined by equation
   (\ref{eq:areal-radius}) above.  As all correction terms are
   scalars, i.e.\ constant on the surface, $H_r$ can only be constant
   for a CE surface.  The term $H - \bar H$ vanishes for a CE surface,
   hence $H_r=0$ only if $R=r$, which is intended.

\section{Numerical issues}
\label{sec:numerics}

   The above definitions are only useful in numerical relativity if
   there is an efficient method to find CE surfaces with a specific
   expansions $\lambda$ or specific areal radii $R$.  This section
   describes such a method, explaining how the surfaces are
   represented and discretised and how equations (\ref{eq:ah-func}),
   (\ref{eq:ce-func}), and (\ref{eq:ar-func}) are then solved.

   Thornburg \cite{ah-finding} gives a rather complete overview of
   current methods of locating apparent horizons, many of which could
   also be applicable for finding surfaces of constant expansion after
   suitable adaptation.  Baumgarte and Shapiro \cite{baumgarte2003}
   review many issues related to numerical relativity, including
   locating apparent horizons.

\subsection{Representation of the surface}

   Surfaces with spherical topology can be represented explicitly by a
   function $h(\theta,\phi)$ that specifies (this is an arbitrary
   choice) the radius $r$ as a function of the spherical coordinates
   $\theta$ and $\phi$.  This choice restricts the possible surfaces
   to those of star-shaped regions about some origin, but this
   restriction does not cause problems in practice; other choices
   would be equally possible.  The surface consists of those points
   $(r,\theta,\phi)$ of the spacelike hypersurface that satisfy the
   condition $r = h(\theta,\phi)$.

   The surface can also be defined implicitly through an expression
   $F(x^i) = 0$, with the level set function $F$ defined e.g.\ as
\begin{equation}
\label{implicit-location}
   F(r,\theta,\phi) := r - h(\theta,\phi) \quad \textrm{.}
\end{equation}
   This representation leads to the equation
\begin{equation}
\label{spacelike-normal}
   s_i = \frac{\nabla_i F}{| \nabla F |}
\end{equation}
   for the outward normal, where $| \nabla F | := \left( \gamma^{jk}\,
   \nabla_j F\, \nabla_k F \right)^\frac{1}{2}$.  The function $F$ and
   the spacelike normal $s^i$ can be calculated using either spherical
   coordinates $(r, \theta, \phi)$, or transformed to use Cartesian
   coordinates $(x, y, z)$.  Such coordinate distinctions are rather
   important for the numerical implementation of an algorithm, so that
   I will not pass over them below.

\subsection{Coordinates on the surface}

   The above choice of spherical coordinates to parameterise the
   surface introduces a preferred coordinate system into the otherwise
   coordinate independent equations (\ref{eq:ah-func}),
   (\ref{eq:ce-func}), and (\ref{eq:ar-func}).  Another preferred
   coordinate system typically comes from the grid used in the time
   evolution code for the whole spacetime.  Often, $\gamma_{ij}$ and
   $K_{ij}$ are given in Cartesian components on a Cartesian grid.
   This raises the question as to which coordinate system to use in
   the numerical implementation of the horizon finder.  The coordinate
   choice will of course not influence the physical results, but it
   can make the implementation more complicated if it requires
   interpolation, or considerably less accurate if it leads to
   coordinate singularities.  It can also make the results more or
   less difficult to interpret, as tensor quantities are numerically
   always expressed through their coordinate components.  Tensors in
   different coordinate systems cannot easily be compared, even if
   they are given at the same grid points.  In short, the choice of a
   coordinate system is numerically an important issue.

   I choose to represent the quantities on the surface using their
   fully three-dimensional Cartesian tensor components, even when they
   only live on the two-dimensional surface.  This has the advantage
   that the tensors $\gamma_{ij}$ and $K_{ij}$ do not have to have
   their components transformed, although they do need to be
   interpolated to the location of the surface.

   The overall advantage of using Cartesian tensor components is that
   coordinate singularities at the poles do not influence the
   representation of such quantities.  While the singularities will
   still influence the calculation of derivatives intrinsic to the
   surface, the representation of the results will not have coordinate
   singularities any more.  Vector and tensor components in spherical
   coordinates would either tend to zero or diverge at the poles.

   I use the indices $i$, $j$, $k$ for Cartesian components ($i,j,k
   \in [x, y, z]$), $u$, $v$, $w$ for spherical components in 3D
   ($u,v,w \in [r, \theta, \phi]$), and $a$, $b$, $c$ for spherical
   components on the surface ($a,b,c \in [\theta, \phi]$).  This
   convention distinguishes which quantities are defined on what
   manifold, and also makes coordinate transformations explicit.

\subsection{Discretisation of the surface}

   The discretisation scheme of the three-dimensional quantities,
   i.e.\ those living on the whole spacelike hypersurface, does not
   play a role when locating a CE surface.  There only has to be a way
   to interpolate these quantities onto the surface.  These quantities
   are often discretised on a Cartesian grid.  It is also possible to
   use spherical coordinates, or to use mesh refinement, without
   influencing the way in which the apparent horizon function $H$ is
   evaluated.

   I choose to discretise the two-dimensional quantities, i.e.\ those
   quantities living on the surface, by using a polar
   $\theta$-$\phi$-grid with constant grid spacings $d\theta$ and
   $d\phi$.  A constant grid spacing works well when the surface has a
   shape that is not too far from a sphere.  Experiments show that
   distorted (peanut-shaped) horizons still work fine, but usually
   require a higher resolution.

\begin{figure}
\begin{center}
{\footnotesize(a)}\includegraphics[scale=0.8]{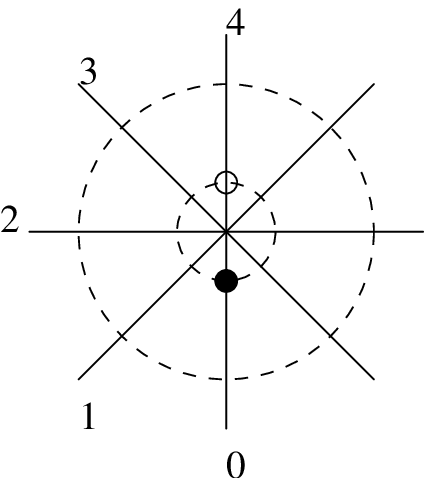}
\hspace{2em}
{\footnotesize(b)}\includegraphics[scale=0.8]{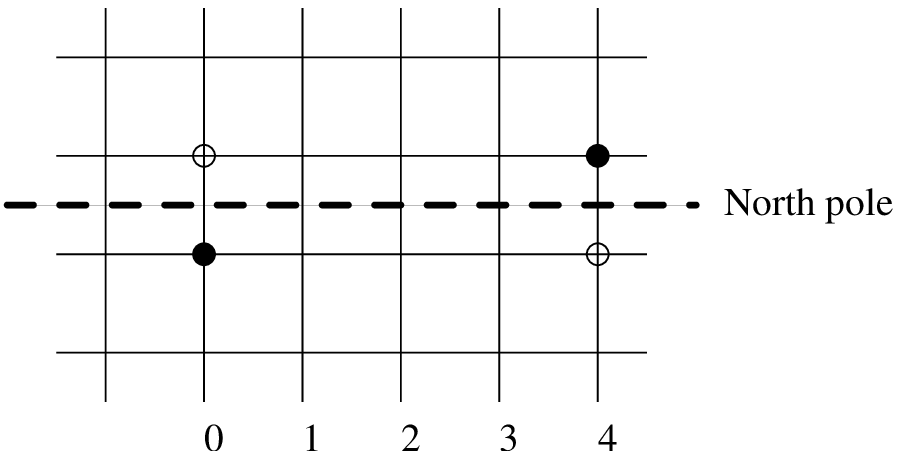}
\end{center}
\caption{
   (a) North pole of the grid on the sphere, in a polar projection.
   This shows the grid lines as they lie on the sphere.
   (b) North pole of the grid on the sphere, as seen from the grid.
   The black and white dots have multiple images on the coordinate
   grid, separated by $\pi$ in the $\phi$-direction.}
\label{theta-boundary}
\end{figure}

   Polar coordinates create coordinate singularities at $\theta=0$ and
   $\theta=\pi$, which I avoid by having the grid points staggered
   with respect to the poles.  The boundary condition in the
   $\phi$-direction is periodic: $f(\theta, \phi) = f(\theta,
   \phi+2\pi)$ for any function $f$ living on the surface.  The
   boundary condition in the $\theta$-direction, i.e.\ across the
   poles, is a bit more involved.  It is $f(\theta, \phi) = P \cdot
   f(-\theta, \phi+\pi)$ for arbitrary tensor components $f$, where
   the parity $P = (-1)^k$ depends on the rank $k$ of the tensor of
   which $f$ is a component.  Figure~\ref{theta-boundary} demonstrates
   the polar boundary condition, especially how the shift by $\pi$ in
   $\phi$-direction comes about.  (See also \cite{shibata2000}.)

   This grid allows numerical partial derivatives to be taken only in
   the $\theta$- and $\phi$-directions.  Equations (\ref{eq:ah-func})
   and (\ref{spacelike-normal}) also need $r$-derivatives, and these
   cannot be taken numerically.  However, the $r$-derivatives of the
   coordinate transformation operators are known analytically, and one
   can see from the definition of $F$ that $\partial_r F = 1$ by
   construction.  Finally, the partial derivatives of the
   three-metric, which are given in Cartesian coordinates, are
   calculated before the metric is interpolated onto the surface.

   My choice of evaluating the apparent horizon function directly on
   the grid that forms the surface while retaining the Cartesian
   coordinate system for the tensor components seems to be an uncommon
   one.  It is also possible to evaluate the apparent horizon function
   not on the surface itself, but instead in the three-dimensional
   spacetime on those grid points that are close to the surface
   \cite{ah-finding, huq-finder, flow-finder}.  This method requires
   interpolating in both directions between the surface and the
   spacelike hypersurface.  It also has the disadvantage that the
   domain of the equation, i.e.\ the set of active grid points,
   changes with the surface, leading to further complications.

   While I represent the shape of the surface using an explicit grid,
   one different way that is commonly used is to expand the surface
   shape in spherical harmonics \cite{nakamura1984, baumgarte1996,
   gundlach1998, anninos1998, alcubierre2000}.  An expansion in
   spherical harmonics does not have coordinate singularities.
   Furthermore, one has direct insight into and control over the high
   spatial frequency components.  A multipole representation is by
   definition restricted to star-shaped regions, while an explicit
   representation is generic; however, this disadvantage is not
   important in practice.  More problematic is that integrations over
   the surface are rather expensive, as they scale with
   $O(\ell_\mathrm{max}^4)$, were $\ell_\mathrm{max}$ is the number of
   multipole moments used.  With an explicit grid, these integrations
   scale with $O(n)$, where $n$ is the number of grid points, and
   where one should choose $n = O(\ell_\mathrm{max}^2)$ for a fair
   comparison.  On the other hand, multipole expansions converge
   faster than finite difference discretisations.  (I present run time
   comparisons in section~\ref{comparison} below.)

   Other possible ways to represent the surface without coordinate
   singularities could use several overlapping grid patches
   \cite{jt-finder}\footnote{It should be pointed out that
   J.~Thornburg, the author of \cite{jt-finder}, and me worked on the
   same problem without knowing from each other (until our works were
   essentially completed).  Incidentally, we also chose very similar
   approaches.}, finite elements, or hierarchical bases
   \cite{yserentant92hierarchical}.

\subsection{Evaluating the CE surface functions}

   I use the algorithm described below to evaluate the apparent
   horizon function $H$.  From that, the CE surface functions
   $H_\lambda$ and $H_r$ can be calculated.

   Although the quantities $F$, $s^i$, and $H$ are defined everywhere
   in space, they only need to be evaluated on the trial surface,
   i.e.\ on the surface determined by $h$.  They still need to be
   defined in a neighbourhood of the surface, so that their
   $r$-derivatives can be taken.  Their continuation off the surface
   is chosen arbitrarily, and is here implicitly determined by the
   above choice of spherical coordinates.  Thus the finder calculates
   quantities on the two-dimensional surface only, although the
   quantities live in three dimensions.

   It would be a bad idea to calculate the derivatives of $s^i$
   numerically, as $s^i$ is itself already calculated as a derivative.
   Taking a derivative of a derivative leads to an effective stencil
   that is twice as large and that contains elements with a weight of
   zero.  These zero weights lead to an even--odd decoupling of the
   grid points, which in turn leads to large solution errors.
   Instead, one uses the chain rule and calculates $\nabla_i s^i$
   directly from second derivatives of $F$.  This makes it necessary
   to calculate $\nabla_i \nabla_j F$ and $\nabla_i s^i$ numerically
   along with $\nabla_i F$ and $s^i$.

\subsection*{Algorithm}

\begin{enumerate}
\renewcommand{\labelenumi}{\arabic{enumi}.}

\item
   Assume that the three-metric $\gamma_{ij}(x^k)$ and the extrinsic
   curvature $K_{ij}(x^k)$ are given as fixed background quantities in
   the whole spacelike hypersurface.  These quantities do not change
   while the CE surface is located.

\item
   Assume that a 2-surface $h(x^a)$ has been chosen for which the
   apparent horizon function is to be evaluated.  This is e.g.\ done
   by an elliptic solver as described below.

\item
   Calculate the surface grid point locations in Cartesian coordinates
   as $x^i(x^a)$, using the definition of $h$.

\item
   Interpolate the three-metric, the extrinsic curvature, and the
   partial derivatives of the three-metric onto the surface.  These
   quantities can be viewed conceptually as representing
   $\gamma_{ij}(x^u)$ and $K_{ij}(x^u)$, but are stored as
   $\gamma_{ij}(x^a)$, $K_{ij}(x^a)$, and $\partial_k
   \gamma_{ij}(x^a)$.  (The partial derivatives are needed later.)
   Note that these tensors now live on the surface, but still have
   their components as Cartesian components.

\item
   Define the implicit surface location $F(x^u) = F(r, \theta, \phi) =
   r - h(\theta, \phi)$.  This quantity is not stored explicitly.

\item
   Calculate an outward-directed covector $\nabla_v F(x^u)$ that is
   orthogonal to the surface.  By definition, $\partial_r F = 1$, and
   $\partial_a F = - \partial_a h$.  This is stored as $\partial_v
   F(x^a)$ and $\partial_v \partial_w F(x^a)$.  (The second
   derivatives are needed later).  The calculation needs the values
   $\partial_b h(x^a)$ and $\partial_b \partial_c h(x^a)$, which are
   calculated on the surface.

\item
   Define the transformation operator $T_i^u$ that transforms from
   spherical to Cartesian covariant coordinate components on the
   surface.  This operator is given by $T_i^u$ = $\partial x^u /
   \partial x^i$, where the $x^i(x^u)$ are given by the usual
   relations
\begin{eqnarray}
\nonumber
   x = r \sin(\theta) \cos(\phi) \quad\quad
   y = r \sin(\theta) \sin(\phi) \quad\quad
   z = r \cos(\theta) \quad\textrm{.}
\end{eqnarray}
   Evaluate this operator and its derivatives at $r=h$, and store them
   as $T_i^u(x^a)$ and $\partial_j T_i^u(x^a)$.

\item
   Transform the components of the outward-directed covector into
   spherical coordinates: $\nabla_i F(x^u) = T_i^v\, \nabla_v F(x^u)$.
   Store $\nabla_i F(x^a)$ and $\nabla_i \nabla_j F(x^a)$.  The
   calculation of the second derivatives needs the derivatives of the
   transformation operator $T$, and also needs the partial derivatives
   of the three-metric for the connection coefficients.

\item
   Normalise the outward-directed covector $\nabla_i F(x^u)$ and raise
   its index to $s^i(x^u) = \gamma^{ij} \nabla_j F(x^u)$.  Calculate
   its derivative $\nabla_i s^i(x^u)$.  This needs the second
   derivatives of $F$.  Store $s^i(x^a)$ and $\nabla_i s^i(x^a)$.

\item
   Calculate the apparent horizon function $H = \nabla_i s^i + K_{ij}
   (s^i s^j - \gamma^{ij})$.  Store it as $H(x^a)$.

\end{enumerate}

   For the above calculation I use a second-order accurate
   interpolator, and second-order centred differences to approximate
   the partial derivatives.  Altogether, the apparent horizon function
   $H$ thus depends on the background spacetime as given by
   $\gamma_{ij}$ and $K_{ij}$, and on the surface location $h$.

\subsection{Solving the CE surface equations}

   I follow the idea presented in \cite{cook1990, cook1992,
   ah-finding, shibata1997, shibata2000, huq-finder} to interpret the
   apparent horizon and CE surface equations as elliptic equations on
   the surface.  Commonly used other methods for apparent horizon
   finding are mean curvature flow \cite{sfb-382-mean-curvature-flow}
   or level flow methods \cite{flow-finder}, or minimisation
   procedures \cite{anninos1998}.  The advantages of flow methods for
   horizon finding are that they are in general more robust in finding
   a horizon, and that they are in practice able to find an
   \emph{outermost}, i.e.\ true apparent horizon, given a large enough
   sphere as initial data.  Their disadvantage is that they lead to
   parabolic or degenerate elliptic equations that are very expensive
   to solve.  Minimisation procedures can also lead to a more stable
   formulation than elliptic formulations, but are usually much
   slower.  I will restrict myself here to the solution of elliptic
   equations.  As described below, I have also implemented a simple
   Jacobi solver that is conceptually identical to a flow method.

   It is possible to change the nature of equations
   (\ref{eq:ah-func}), (\ref{eq:ce-func}), and (\ref{eq:ar-func}) by
   extending them to the three-dimensional spacelike hypersurface.
   This is done by changing from the unknown function $h(\theta,\phi)$
   to a level set $h_q(\theta,\phi)$, where $q$ is a level set
   parameter \cite{sfb-382-mean-curvature-flow}.  The advantage of
   such an extension is that multiple CE surfaces can be found at the
   same time.  The disadvantage is that such an extension has a higher
   numerical cost because of the additional dimension.  However, such
   a method would be ideal to find the whole set of CE surfaces that
   make up a foliation at once.  Unfortunately, one would have to know
   in advance what region of the hypersurface is not covered by the
   foliation, and impose a corresponding inner boundary condition.

   The conditions $H=0$, $H_\lambda=0$, and $H_r=0$ are all nonlinear
   elliptic equations in $h$, as can be seen by substituting the
   definitions of $s^i$ and $F$ into $H$.  Strictly speaking, only the
   equations $| \nabla F |\, H = 0$ etc.\ are elliptic.  But this
   additional factor is often close to unity, and omitting it makes
   then little difference for nonlinear elliptic solvers.

   In order to solve a nonlinear equation, one needs initial data,
   which are in this case an initial guess for the surface location.
   These initial data select which CE surface will be found --- if
   there are several present on the spacelike hypersurface --- or
   whether any will be found at all.  During a time evolution one can
   easily use the location from the previous time step as initial
   data; this works very well in practice and is called
   \emph{tracking}.  However, finding CE surfaces in an unknown
   spacetime is more difficult.  Depending on how e.g.\ the initial
   configuration for a time evolution run is constructed, one does not
   know where or of what shape the CE surfaces are.

   I have implemented this CE surface finding algorithm within the
   Cactus framework \cite{cactus-home-page}, which has gained some
   popularity in the numerical relativity community.  I hope that this
   will make it easier for other people to use this implementation not
   only as an external program, but also directly from within their
   code.  Within Cactus, I have examined two different nonlinear
   elliptic solvers for the CE surface equations, a Newton-like method
   and a flow-like method.

\subsubsection{Newton method}
\label{sec:newton-method}

   The faster of the two solvers uses a Newton-like method to reduce
   the nonlinear equation to a linear one, and then a standard solver
   for the linear problem (GMRES with ILU preconditioning).  This is
   implemented by interfacing to the PETSc library
   \cite{petsc-home-page, petsc-manual}.  While this solver is very
   fast (it usually takes less than a second to converge), it also
   converges only if one starts sufficiently close to the final
   location and shape.  In particular, it has problems with finding
   distorted surfaces when starting with a sphere and vice versa.

   PETSc needs the Jacobian of the elliptic equation in order to solve
   the system.  I chose to represent the Jacobian explicitly as a
   sparse matrix.  It is then necessary to have a routine that creates
   the structure of the non-zero elements of this Jacobian.  In doing
   so one has to make sure that the grid point couplings due to the
   boundary conditions (see figure~\ref{theta-boundary}) are treated
   correctly.  PETSc can then calculate the values of the non-zero
   entries through numeric differentiation, which is usually slow, but
   is in this case fast enough.
   Although PETSc also offers matrix-free methods, i.e.\ solvers that
   only use implicit representations of the matrix, I did not use
   them, because they can lead to convergence problems that are
   difficult to investigate.

   This solver does not always work when solving for CE surfaces with
   a specified areal radius $R$.  I assume that this is because I
   only use an approximation to the Jacobian in this case.  The term
   $\bar H$ in equation (\ref{eq:ar-func}) leads to a fully populated
   Jacobian, whereas I ignore the $\bar H$ term when I calculate it.

   One large advantage of PETSc is that it can easily be reconfigured
   through options at run time to use a variety of different solvers
   and preconditioners.  It can also explicitly calculate all
   eigenvalues of a system (if the system is small), which is very
   helpful for understanding convergence properties.

\subsubsection{Flow method}

   The other, more robust solver is a Jacobi solver, which is
   conceptually identical to a mean curvature flow solver, as e.g.\
   described in \cite{tod1991, flow-finder}.  (It should probably
   rather be called an expansion flow solver, as the flow really is
   governed by the surface's expansion and not its mean curvature.)
   This solver converges very reliably to the desired CE surface if
   the step size is chosen to be small enough.  A Jacobi solver solves
   in effect a parabolic equation, so that the maximum stable step
   size scales with the square of the surface resolution.  This
   requires very small step sizes unless the surfaces resolution is
   very coarse.  That means that this solver requires very long run
   times of the order of minutes, or tens of minutes if one starts far
   away from the solution.

   The maximum stable step size depends on the equation's eigenvalues,
   which are unknown because the equation is nonlinear.  I have
   therefore implemented a primitive adaptive step size control
   mechanism which monitors the norm of the residual.  As long as the
   residual decreases while iterating, the step size is slowly
   increased (e.g.\ by 10\% per step).  If the residual increases, the
   step size is sharply decreased (e.g.\ by a factor of 10).  I find
   that it is not necessary to undo steps that increase the residual;
   on the contrary, such steps usually increase the convergence rate.
   I assume that such overshooting gives this algorithm some SOR-like
   properties.  This mechanism, which surely could be improved upon,
   chooses step sizes which usually lead to a stable integration,
   while it also adapts to the stability criteria that change with the
   flowing surface.

   Of course, even adaptive step size control does not lead to a truly
   fast algorithm for a parabolic equation.  In order to accelerate
   this solver, I often use a hybrid approach: I start by using the
   flow solver, but do not wait until it has fully converged.  After
   some time I switch to using PETSc, with the output of the flow
   solver as the starting point.  This combines the robustness of the
   flow solver in finding unknown shapes with the fast convergence
   rate of PETSc once an approximate solution is known.

\section{Tests}
\label{sec:tests}

\subsection{Tests with analytic solutions}

   As first test for the convergence properties of this method, I use
   a single black hole in Kerr--Schild coordinates (see e.g.\
   section~3.3.1 of \cite{livingreviews-cook} for a definition).
   Table~\ref{convtest-single-1-results} contains results from runs
   with a black hole of mass $M=1$ and spin $a=0$.  The table compares
   the irreducible mass of the horizon with the ADM mass of the whole
   spacelike hypersurface which I calculate numerically for reference
   (see e.g.\ equation (7.6.22) in \cite{adm-mass}).  I would consider
   the resolutions $1/4$, $1/8$, and $1/16$ to be coarse, reasonable,
   and fine, respectively, for a numerical representation of this
   spacetime.

\begin{table}
\caption{
   Left: Horizon and ADM masses for a black hole with $M=1$ and $a=0$.
   $R/2$ is the irreducible mass of the horizon where $A = 4\pi R^2$
   is its area, and $M_\mathrm{ADM}$ is the ADM mass of the whole
   spacelike hypersurface.  Both the spacetime and the surface vary in
   resolution.  The last row gives the analytic values.
   Right: Convergence factors for these results.  A value of 4
   indicates second order convergence.}
\label{convtest-single-1-results}
\label{convtest-single-1-cfacts}
\begin{indented}
\item[]
\begin{tabular}{@{}lllll}\br
run & $dx$   & $d\phi$    & $R/2$    & $M_\mathrm{ADM}$ \\\mr
a   & $1/4$  & $2\pi/36$  & 0.991111 & 1.00043          \\
b   & $1/8$  & $2\pi/72$  & 0.997693 & 1.00010          \\
c   & $1/16$ & $2\pi/144$ & 0.999429 & 1.00002          \\
    & $0$    & $0$	  & 1.0	     & 1.0              \\\br
\end{tabular}
\hspace{2em}
\begin{tabular}{@{}lll}\br
runs	& $R/2$	& $M_\mathrm{ADM}$ \\\mr
a, b, c	& 3.79	& 4.43		   \\
a, b	& 3.85	& 4.37		   \\
b, c	& 4.04	& 4.18		   \\\br
\end{tabular}
\end{indented}
\end{table}

   The same table also shows the convergence factors for three-way and
   two-way convergence tests from the above runs.  The convergence
   factors $f$ are calculated in the usual way via $f =
   (v_1-v_2)/(v_2-v_3)$ or $f = (v_1-v_a)/(v_2-v_a)$, where $v_1$,
   $v_2$, $v_3$, and $v_a$ represent the coarse, medium, fine grid,
   and analytic values, respectively.  A factor of 4 indicates
   second-order convergence.  The results demonstrate that a spatial
   resolution of $dx=1/4$ is not sufficient to be in the convergent
   regime, and also that the numerical ADM masses reported here might
   not be very accurate.

   In another convergence test, shown in
   table~\ref{convtest-single-2-results}, I keep the spatial
   resolution of the spacelike hypersurface constant at $dx=1/8$, and
   vary only the resolution of the surface grid.  This time the black
   hole has a mass $M=1$ and a spin $a_z=1/2$, again in Kerr--Schild
   coordinates.  The apparent horizon spin $a_z$ is estimated via its
   equatorial circumference as described below.  The last line of the
   table shows the analytic values.

\begin{table}
\caption{
   Left: Horizon masses, spins, and ADM masses for a black hole with
   $M=1$ and $a_z=1/2$.  Only the resolution of the surface varies.
   The last row gives the analytic values.
   Right: Convergence factors for these results.  A value of 4
   indicates second order convergence.}
\label{convtest-single-2-results}
\label{convtest-single-2-cfacts}
\begin{indented}
\item[]
\begin{tabular}{@{}lllll}\br
run	& $d\phi$	& $R/2$		& $a_z$		& $M$      \\\mr
A	& $2\pi/36$	& 0.960886	& 0.496507	& 0.994655 \\
B	& $2\pi/72$	& 0.964675	& 0.498183	& 0.998537 \\
C	& $2\pi/144$	& 0.965594	& 0.498815	& 0.999511 \\
D	& $2\pi/288$	& 0.965819	& 0.498997	& 0.999754 \\
	& 0		& 0.965926	& 0.5		& 1.0	   \\\br
\end{tabular}
\hspace{2em}
\begin{tabular}{@{}llll}\br
runs	& $R/2$	& $a_z$	& $M$  \\\mr
A, B, C	& 4.13	& 2.65	& 3.99 \\
B, C, D	& 4.08	& 3.47	& 4.01 \\\br
\end{tabular}
\end{indented}
\end{table}

   Because it is known in this case that the apparent horizon has a
   spin that is aligned with the $z$ axis and that the horizon is not
   distorted, one can use the area and the equatorial circumference to
   calculate the spin magnitude and the total mass.  In the elliptic
   coordinates that one prefers for Kerr black holes (see e.g.\
   section~3.3 of \cite{livingreviews-cook}), a horizon with mass $M$
   and spin $a$ is located at the coordinate radius $r$, has the area
   $A$ and the equatorial circumference $L$, which are given by:
\begin{eqnarray}
\label{eq:bh-spin-estimate}
   r = M + \sqrt{M^2 - a^2} \quad\quad
   A = 4\pi\, (r^2 + a^2)   \quad\quad
   L = 2\pi\, (r^2 + a^2)/r
\end{eqnarray}
   (I thank Badri Krishnan for pointing out the last equation to me.)
   These equations can be solved for $a$ and $M$.

   The convergence factors for the above test are also shown in
   table~\ref{convtest-single-2-cfacts}.  As the spacetime is now
   fixed, and differs slightly from the analytic solution due to the
   discretisation errors, I omit the two-way convergence tests here.
   While the estimate for the total mass $M$ that is calculated this
   way has a reasonable accuracy, the spin estimate does not converge
   to second order for these resolutions.\footnote{Equations
   (\ref{eq:bh-spin-estimate}) contain only the term $a^2$.  When the
   spin is small, the estimate for $a^2$ can even be negative.  This
   is mainly caused by accumulation of numerical errors: An
   $O(10^{-2})$ error in $a^2$ shows up as an $O(10^{-1})$ error in
   $a$.}  This is unfortunate, but not of much consequence, as this
   method for calculating the spin is not applicable in general
   anyway.

\subsection{Comparison to other apparent horizon finders}
\label{comparison}

   I compare the accuracy and speed of the elliptic method described
   in this article to two other commonly used apparent horizon finding
   methods, namely the \emph{fast flow} algorithm presented in
   \cite{gundlach1998} and a minimisation procedure described in
   \cite{anninos1998}.  Implementations for both are publicly
   available with the Cactus code \cite{cactus-home-page} and have
   been described in \cite{alcubierre2000}.  Both expand the apparent
   horizon surface in spherical harmonics.  They are part of the
   Cactus infrastructure, are used in production runs, and have been
   thoroughly optimised for speed.

   As test bed I use a black hole in Kerr--Schild coordinates with
   $M=1$ and $a_z=0.6$, discretised on a Cartesian grid with $dx=1/8$.
   The horizon is then located at $r = M+\sqrt{M^2-a^2} = 1.8$, which
   corresponds to the coordinate location $x^2+y^2+z^2 = \rho^2 =
   r^2+a^2 (1-z^2/r^2) = 3.6-z^2/9$.  In these coordinates, the
   apparent horizon is thus a slightly oblate ellipsoid; the polar
   coordinate radius is $\rho_\mathrm{po} \approx 1.868$ and the
   equatorial coordinate radius is $\rho_\mathrm{eq} \approx 1.897$.
   The apparent horizon has octant symmetry (although the spacetime
   does not), and I enforce it explicitly.

   I set up two test cases for these data.  In the first case I start
   with a distorted, oblate surface, and let the finders converge to
   the horizon.  The initial surface has a polar radius of
   $\rho_\mathrm{po} = 2$ and an equatorial radius of
   $\rho_\mathrm{eq} = 3.5$, or $\ell_0 \approx 3$ and $\ell_2 \approx
   -0.45$ in spherical harmonics (with the normalisations as described
   in \cite{alcubierre2000}).  This is supposed to mimic a situation
   where not much is known about the spacetime, so that the initial
   guess is far from the solution.  In the second test case I start
   with a sphere at $\rho=2$, i.e.\ close in location and shape to the
   actual horizon.  This allows the use of much faster algorithms.
   This situation is similar to horizon (or CE surface) tracking.

   I use the three finders each with two different configurations.
   For the elliptic method I choose the configurations depending on
   the test case.  For the first test case, I first use the Jacobi
   solver, and then follow up by calling PETSc, as described in
   section~\ref{sec:newton-method} above.  For the second test case
   that corresponds to horizon tracking, I use only PETSc.  For the
   \emph{fast flow} method and the minimisation procedure I use once
   the default configuration suggested by the authors of the
   implementation, and once an adapted configuration with more
   multipole moments and a coarser spatial resolution of the surface.

   The numerical resolution of the spacelike hypersurface leads to
   about $4\pi\rho^2\, dx^2 \approx 2800$ grid points ``on'' the
   horizon, i.e.\ about $350$ on the octant that is represented
   numerically.  For the explicit surface representation in the
   elliptic method I use $12^2=144$ grid points.  For the adapted
   configuration of the \emph{fast flow} method and the minimisation
   procedure, I attempt to choose settings for the spherical harmonics
   that are equivalent to the settings for the explicit surface
   representation.  I decided to limit the expansion to
   $\ell_\mathrm{max} = 12$, which leads to $28$ nonvanishing
   spherical harmonic coefficients due to the octant symmetry.  This
   expansion still has a lower spatial resolution than $12^2=144$
   explicit grid points, but spherical harmonics have better
   approximation properties, as can also be seen from the errors in
   the result.  The number $\ell_\mathrm{max}=12$ is suggested as
   ``high enough'' in the documentation of the code.  These adapted
   settings should also accelerate the \emph{fast flow} method and the
   minimisation procedure, leading to a fairer comparison of the run
   times.

   Table~\ref{tab:comparison-configuration} lists the exact parameters
   used for the finders.  Table~\ref{tab:comparison-physics} compares
   the physical results, i.e.\ area, equatorial circumference, mass,
   and spin as calculated via equations (\ref{eq:bh-spin-estimate}).
   The methods using a decomposition in spherical harmonics apparently
   slightly depend on the initial data.  They also have a higher
   accuracy.  Table~\ref{tab:comparison-speed} lists the run times
   that the finders took on a Pentium III processor with 1200 MHz,
   i.e.\ on reasonable hardware in 2003.  The elliptic method with the
   explicit surface representation is more than an order of magnitude
   faster at horizon tracking.  When the resolution of the \emph{fast
   flow} method is reduced down to $\ell_\mathrm{max}=4$, then the run
   time diminishes to $4.10$ seconds, which is still considerably
   larger than for the explicit representation.  Run times are
   important during a time evolution run, because a horizon finder has
   to run at every (or every few) time step(s).  In order to determine
   a CE surface foliation, the finder even has to run many times for
   each hypersurface.

\begin{table}
\caption{
   Configurations of the horizon finders.  $\ell_\mathrm{max}$ is the
   highest term in the spherical harmonic expansion.  $n_\theta$ and
   $n_\phi$ are the number of grid points on the surface.  (The
   \emph{fast flow} solver and the minimisation procedure need an
   explicit surface for surface integrals.)}
\label{tab:comparison-configuration}
\begin{indented}
\item[]
\begin{tabular}[t]{@{}lrr}\br
\multicolumn{3}{@{}l}{minimisation}       \\\mr
                    & default & adapted \\
$\ell_\mathrm{max}$ &     $8$ &    $12$ \\
$n_\theta$          &   $100$ &    $12$ \\
$n_\phi$            &   $100$ &    $12$ \\\br
\end{tabular}
\hspace{2em}
\begin{tabular}[t]{@{}lrr}\br
\multicolumn{3}{@{}l}{\emph{fast flow}}   \\\mr
                    & default & adapted \\
$\ell_\mathrm{max}$ &     $8$ &    $12$ \\
$n_\theta$          &   $100$ &    $12$ \\
$n_\phi$            &   $100$ &    $12$ \\\br
\end{tabular}
\vspace{2ex}

\item[]
\begin{tabular}[t]{@{}lr}\br
\multicolumn{2}{@{}l}{elliptic/flow} \\\mr
$n_\theta$      & $12\quad$       \\
$n_\phi$        & $12\quad$       \\
flow iterations & $20000\quad$    \\
flow step size  & $10^{-4}$       \\
error tolerance & $10^{-8}$       \\\br
\end{tabular}
\hspace{2em}
\begin{tabular}[t]{@{}lr}\br
\multicolumn{2}{@{}l}{elliptic/PETSc} \\\mr
$n_\theta$      & $12\quad$        \\
$n_\phi$        & $12\quad$        \\
error tolerance & $10^{-8}$        \\\br
\end{tabular}
\end{indented}
\end{table}

\begin{table}
\caption{
   Results for different configurations and horizon finder algorithms.
   As expected, the expansion in spherical harmonics leads to a better
   representation of the surface.  Note that the analytic values in
   the last row are not exact, as they do not take the discretisation
   of the spacelike hypersurface into account.}
\label{tab:comparison-physics}
\begin{indented}
\item[]
\begin{tabular}{@{}llllllll}\br
configuration & finder                     &    $A$ &    $L$ &   $R/2$ &   $a_z$ &     $M$ \\\mr
1 (finding)   & minim.\ (default)          & 45.146 & 12.554 & 0.94770 & 0.59962 & 0.99902 \\
1 (finding)   & minim.\ (adapted)          & 45.086 & 12.554 & 0.94710 & 0.60279 & 0.99902 \\
1 (finding)   & elliptic (flow)            & 44.843 & 12.513 & 0.94452 & 0.59803 & 0.99574 \\\ms
2 (tracking)  & \emph{fast flow} (default) & 45.224 & 12.562 & 0.94852 & 0.59891 & 0.99965 \\
2 (tracking)  & \emph{fast flow} (adapted) & 45.131 & 12.558 & 0.94755 & 0.60212 & 0.99933 \\
2 (tracking)  & elliptic (PETSc)           & 44.843 & 12.513 & 0.94452 & 0.59803 & 0.99574 \\\ms
\multicolumn{2}{@{}l}{reference values}    & 45.239 & 12.566 & 0.94868 & 0.6     & 1.0     \\\br
\end{tabular}
\end{indented}
\end{table}

\begin{table}
\caption{
   Run times for different configurations and horizon finder
   algorithms.  The elliptic method with the explicit surface
   representation is very fast at horizon tracking.}
\label{tab:comparison-speed}
\begin{indented}
\item[]
\begin{tabular}{@{}llr}\br
configuration & finder                     & time [s] \\\mr
1 (finding)   & minim.\ (default)          &  98.94   \\
1 (finding)   & minim.\ (adapted)          &  91.45   \\
1 (finding)   & elliptic (flow)            & 190.43   \\\ms
2 (tracking)  & \emph{fast flow} (default) &  26.97   \\
2 (tracking)  & \emph{fast flow} (adapted) &  10.82   \\
2 (tracking)  & elliptic (PETSc)           &   0.50   \\\br
\end{tabular}
\end{indented}
\end{table}

\section{Applications of CE surfaces}
\label{sec:applications}

\subsection{Comparing different coordinate systems}

   Empirically, surfaces of constant expansion seem indeed to foliate
   spacelike hypersurfaces.  Assuming that this is true in general,
   they can be used to compare spacetime metrics that are given in
   different coordinate systems.  As CE surfaces and thus the
   foliations depend on the slicing, this requires that equivalent
   slicings are obtained in some way.  That is in general a difficult
   problem, and I only want to mention two simple ways here.  One
   trivial way is to restrict oneself to considering time-symmetric
   initial data, where one can compare the results of varying
   parameters such as black hole configurations or the numerical
   resolution.  Another way is to perform a time evolution, starting
   with identical initial data and using a maximal slicing condition,
   comparing results e.g.\ obtained with different shift conditions or
   with different numerical parameters.

   In order to demonstrate comparing different metrics, I use in the
   following an artificial setup with known properties.  I compare two
   metrics of identical stationary Kerr--Schild black hole spacetimes,
   where a coordinate transformation has been applied to the second
   metric.  I choose the coordinate transformation so that it changes
   the three-metric only and does not modify the slicing of the
   spacetime.  The different three-metrics lead to different extrinsic
   curvatures as functions of the three-coordinates.

   I use Kerr--Schild black holes with $M=1$ and $a_z=0.6$.  The
   apparent horizon in this spacetime is also an isolated horizon.
   The transformation is a compression along the $z$ axis by a factor
   of $0.5$, and a shear in the $x$-$z$ plane by a factor of $0.5$.
   In this case, the spacetime is deformed by a single, global, linear
   transformation, so that the spin can still be described by a
   direction vector.

\begin{figure}
\begin{center}
\includegraphics[scale=1.1]{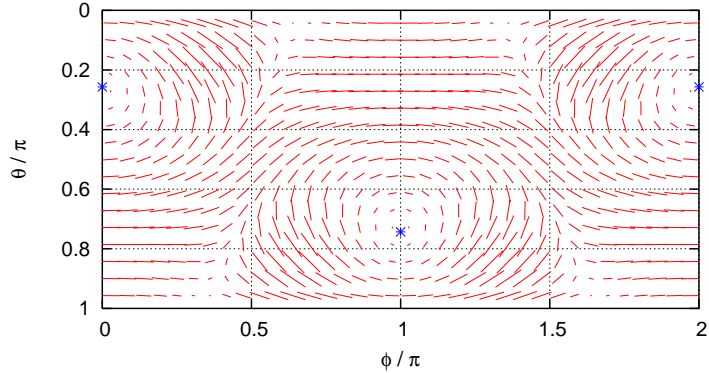}
\end{center}
\caption{
   Killing vector field $\xi^a$ on the apparent horizon in the
   $\theta$-$\phi$ plane.  The $z$ axis penetrates this plane at
   $\theta=0$ and $\theta=\pi$, the positive $x$ axis at
   $\theta=\pi/2$, $\phi=0$.  The two singular points where $\xi^a=0$
   (marked by the stars) are the black hole's ``poles'' defined by the
   spin.}
\label{fig:ksd-spin}
\end{figure}

   I calculate the Killing vector field on this horizon in the way
   described in \cite{dreyer2003}.  This Killing vector field, shown
   in figure~\ref{fig:ksd-spin}, defines the direction of the spin,
   putting the ``poles'' of the spin at $(\theta,\phi) = (0.257\pi,0)$
   and $(0.743\pi,\pi)$, which is close to the real values $(\pi/4,0)$
   and $(3\pi/4,\pi)$.  From the Killing vector field also follow a
   spin magnitude of $a=0.601$ and a total mass of $M=0.997$, which
   are also close to the real values.

\begin{figure}
\begin{center}
{\footnotesize(a)}\includegraphics{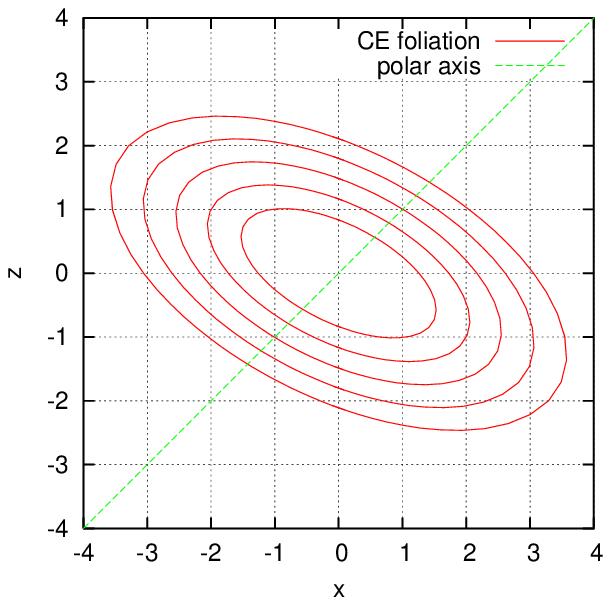}
\hspace{2em}
{\footnotesize(b)}\includegraphics{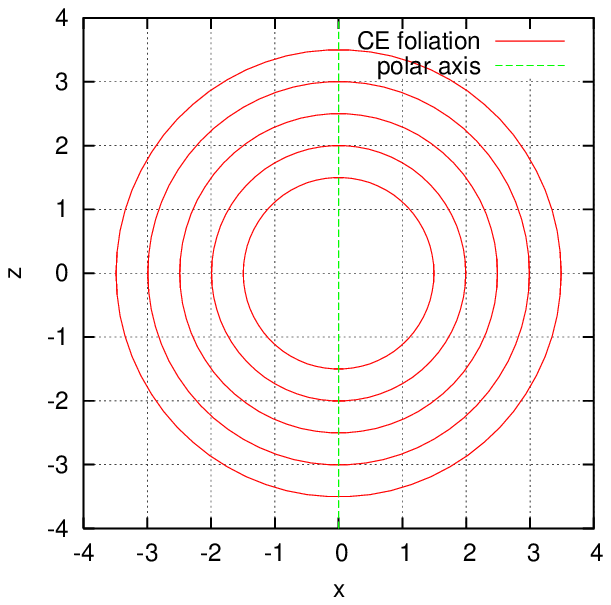}
\end{center}
\caption{
   (a) CE foliation of the transformed Kerr--Schild black hole with $R
   = \frac{3}{2}, \frac{4}{2} \cdots \frac{7}{2}$ (from inner to outer
   surface).  The $R=2$ surface is the apparent horizon.  The axis of
   the black hole is defined by its spin, as shown in
   figure~\ref{fig:ksd-spin}.
   (b) CE foliation of the undistorted Kerr--Schild black hole for
   comparison for the same values of $R$.  The $R=2$ surface is again
   the apparent horizon.}
\label{fig:ksd}
\end{figure}

   Figure~\ref{fig:ksd} shows a cut in the $x$-$z$ plane through the
   numerically located CE surfaces in the transformed metric, where
   the direction of the spin is also indicated.  It also shows the
   same cut through the undistorted black hole for comparison.  The
   Killing vector field and the horizon area could be used to define
   invariant lines of latitude and longitude, providing a complete
   invariant coordinate system within the spacelike hypersurface.
   Section~VI of \cite{dreyer2003} discusses a different and
   slicing-independent method to obtain such a coordinate system.

\subsection{Detecting coordinate distortions}

   Surfaces of constant expansion are useful even when there is no
   apparent horizon present, where they can be used to detect
   distorted coordinate systems.  Especially when the spacetime is the
   result of a time evolution, it is difficult to judge how much of an
   observed distortion is caused by the gauge condition.  Surfaces of
   constant expansion can here serve as anchors, because they are
   defined independently of the gauge choice that is contained in the
   three-metric.

   CE surfaces can be considered to be a generalisation of spheres in
   Euclidean geometry.  (In the case of vanishing extrinsic curvature,
   CE surfaces have constant mean curvature, which is a defining
   criterion of a sphere.)  It is thus justified to define an adapted
   ``undistorted'' coordinate system by requiring that CE surfaces be
   coordinate spheres in it.  A related condition, namely that
   constant mean curvature surfaces be coordinate spheres, has been
   proposed as a condition for the radial shift vector during time
   evolution \cite{krivan1995}.  Invariant lines of latitude and
   longitude could be obtained by defining them at spatial infinity
   and propagating them inwards perpendicular to the CE surfaces.

   I consider here an axisymmetric, subcritical Brill wave
   \cite{brill} as example.  The three-metric can be written as
\begin{eqnarray}
   ds^2 = \psi^4 \left[ e^{2q} (d\rho^2 + dz^2) + \rho^2 d\phi^2
   \right]
\end{eqnarray}
   with the free function $q(\rho,z)$.  I use the Holz source function
\begin{eqnarray}
   q = A\, \rho^2\, e^{-r^2}
\end{eqnarray}
   with the amplitude parameter $A$.  $\rho$ is here the cylindrical
   radial coordinate with $\rho^2 = x^2 + y^2$.  The initial data are
   chosen to be time-symmetric.  Constructing these data involves a
   numerical solution of the Hamiltonian constraint.

\begin{figure}
\begin{center}
\includegraphics[scale=1.1]{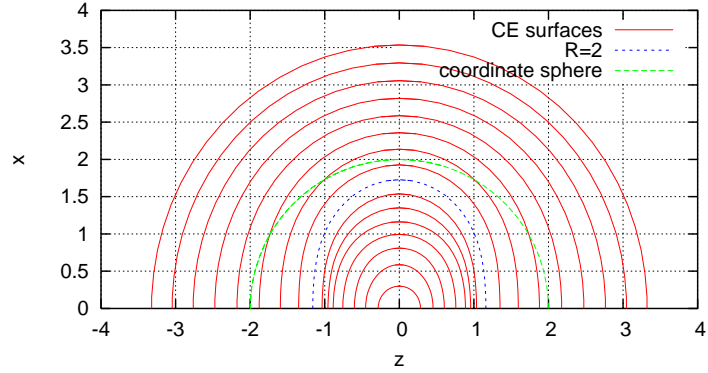}
\end{center}
\caption{
   CE foliation of an axisymmetric Brill wave.  Displayed is a cut
   through the $x$-$z$ plane.  (Note the directions of the coordinate
   axes.)  The CE surfaces have areal radii $R = \frac{1}{4} \ldots 4$
   in steps of $\frac{1}{4}$; the surface with $R=2$ is emphasised.
   Note the coordinate distortion of the CE surfaces; a coordinate
   sphere with $r=2$ is drawn for comparison.  There is no apparent
   horizon present.}
\label{brill-wave}
\end{figure}

\begin{figure}
\begin{center}
\includegraphics{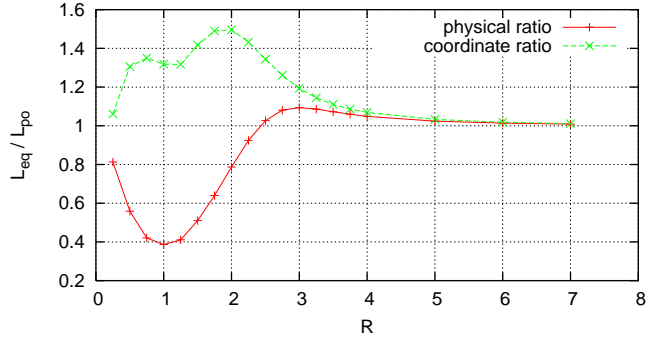}
\end{center}
\caption{
   Ratio between the equatorial and polar physical circumferences for
   the Brill wave CE surfaces.  Values smaller than $1$ indicate a
   prolate surface, values larger than $1$ an oblate surface.  For
   comparison the ratio of the corresponding coordinate circumferences
   is also shown.}
\label{brill-wave-cf}
\end{figure}

   I choose the amplitude $A=4$ and calculate a CE foliation of these
   data, shown in figure~\ref{brill-wave}.  The CE surfaces near $R=2$
   are quite far from coordinate spheres.  This distortion is not an
   intrinsic property of the gravitational wave content; rather, it is
   only a property of this particular coordinate system.
   Figure~\ref{brill-wave-cf} compares the equatorial and polar
   circumferences of the CE surfaces, showing that these surfaces are
   prolate, although they have an oblate coordinate shape in the Brill
   coordinate system.  The same shape parameter, albeit for event
   horizons, is used in \cite{event-horizons} to detect quasinormal
   modes of black holes.

\subsection{CE surfaces for multiple black holes}

   In order to prepare and analyse initial data for binary black hole
   collision runs, I superpose two Kerr--Schild black holes as
   proposed in \cite{superposed-kerrschild}, and then solve the
   constraint equations.  This results in spacetimes with either two
   separate or a single distorted (``merged'') black hole(s).  One can
   freely specify the locations, momenta, masses, and spins of the
   black holes prior to the superposition.  The relation to the
   properties of the superposed black holes is not known, but it is
   hoped that the superposition will not change them by much.

   For the superposition, I choose a grid spacing of $dx=1/4$, an
   outer boundary that has a distance $8$ from the centres of the
   black holes, and excise a spherical region with radius $1$ around
   the singularities.  I superpose the three-metric and the extrinsic
   curvature without attenuation.  I then solve the constraint
   equations numerically using the York-Lichnerowicz method with a
   conformal transverse-traceless decomposition.  (This method is
   described e.g.\ in section~2.2.1 of \cite{livingreviews-cook}.)  At
   the outer boundary I use a Dirichlet condition for the vector
   potential and a Robin condition for the conformal factor.  At the
   excision boundaries I use Dirichlet conditions, keeping the
   boundary values from the superposition.

   Whether or not a common apparent horizon exists is a hint as to
   whether the black hole configuration contains two separate or a
   single merged black hole(s); this fact is not known otherwise.  The
   only way to find out would be to evolve this configuration for a
   long enough time so that the event horizon could be tracked
   backwards in time.  This is unfortunately not easily possible (if
   at all) with today's black hole evolution methods.

\begin{figure}
\begin{center}
\includegraphics{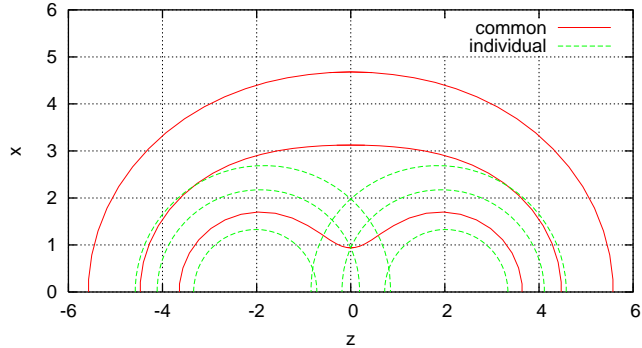}
\end{center}
\caption{
   Common and individual CE surfaces for superposed Kerr--Schild black
   holes with singularities at $z=\pm 2.0$.  (Note the directions of
   the coordinate axes.)  The expansions and areal radii are compared
   in table~\ref{tab:ksc2}.  The middle common CE surface is the
   outermost apparent horizon.  The middle individual CE surfaces are
   also apparent horizons.}
\label{fig:ksc2}
\end{figure}

   Figure~\ref{fig:ksc2} compares common and individual CE surfaces
   for two superposed Kerr--Schild black holes with $M=1$ and $a=0$ at
   $z=\pm 2.0$.  This configuration is axisymmetric about the $z$ axis
   and has a reflection symmetry about the $x$-$y$ plane.  The system
   has a common apparent horizon.  The individual CE surfaces do not
   extent far outwards; they cease to exist for values of $R \gtrsim
   3$.  Similarly, the common CE surfaces do not extend arbitrarily
   far inwards, ceasing to exist for $R \lesssim 2.7$.  It is also
   clear that there is no smooth transition between the two
   foliations.  However, there is an overlap between the regions in
   which the different foliations are valid.  In both foliations the
   area increases monotonically outwards, as shown in
   table~\ref{tab:ksc2}.

\begin{table}
\caption{
   Expansions and areal radii for the CE surfaces shown in
   figure~\ref{fig:ksc2}.  The largest individual CE surface shown has
   a larger expansion (and areal radius) than the smallest shown
   common CE surface.}
\label{tab:ksc2}
\begin{indented}
\item[]
\begin{tabular}{@{}ll}\br
\multicolumn{2}{@{}c}{common} \\
$\;\;\;\lambda$ & $R$         \\\mr
$+0.1$          & $5.053$     \\
$\;\;\;0.0$     & $3.804$     \\
$-0.25$         & $2.713$     \\\br
\end{tabular}
\hspace{2em}
\begin{tabular}{@{}ll}\br
\multicolumn{2}{@{}c}{individual} \\
$\;\;\;\lambda$ & $R$             \\\mr
$+0.1$          & $2.952$         \\
$\;\;\;0.0$     & $2.389$         \\
$-0.5$          & $1.494$         \\\br
\end{tabular}
\end{indented}
\end{table}

\subsection{Apparent horizon pre-tracking}
\label{sec:pre-tracking}

   During a time evolution simulation, new apparent horizons may
   appear.  One does in general not know their location and not even
   their time of appearance in advance.  The new apparent horizon can
   be formed in a collapse, or it can be the common horizon in a
   binary black hole coalescence.  \cite{flow-finder} describes a
   mechanism in an apparent horizon finder that automatically detects
   whether a two black hole system has a common horizon, and if not,
   finds the two separate horizons instead.  However, this flow finder
   is slow, and it is not feasible to apply it at every time step
   during a time evolution.  Except for choosing various trial
   surfaces as initial guesses and testing whether one of them
   converges to an apparent horizon, there is no other way to find out
   whether a new apparent horizon has appeared.  This is a slow and
   cumbersome method.

   Surfaces of constant expansion offer a more efficient approach
   here.  Apparent horizon pre-tracking is an algorithm by which a CE
   surface with decreasing expansion $\lambda$ is tracked during a
   binary black hole time evolution before a common apparent horizon
   has formed.  It is much faster to track a surface that is a
   candidate for an apparent horizon than to examine the spacelike
   hypersurfaces anew at each time step.

   Pre-tracking works as follows.  One starts with a CE surface that
   ``tightly'' encloses both singularities.  If this surface is
   sufficiently large, then it will have a positive expansion
   $\lambda$.  This surface is tracked in time.  At each time step of
   the evolution, one iteratively tries to find a surface with as
   small an expansion as possible.  In practice, this minimum possible
   expansion will vary only slowly with time.  Finding a surface with
   a smaller expansion is very fast, because a surface with a larger
   expansion is already known and can serve as an initial guess.  As
   soon as a surface with an expansion of zero can been found, a new
   apparent horizon has been detected.

   This method allows the locating of new apparent horizons as soon as
   they appear.  The current value of $\lambda$ is also an indication
   as to how close the spacelike hypersurface is to having an apparent
   horizon near the CE surface.

   As an example I consider a system of two Brill--Lindquist black
   holes.  I do not want to go into the problems associated with the
   time evolution of a two black hole system, where one does not know
   when and where the common apparent horizon is supposed to appear.
   Instead I consider a pseudo-evolution consisting of a series of
   Brill--Lindquist black holes with decreasing distances.  While this
   series does not form a proper time evolution, it is still a useful
   example demonstrating pre-tracking.

   In this series the singularities are located at $\pm z$ on the $z$
   axis.  I start the series with $z=1.00$ and ``evolve'' it in steps
   of $\Delta z=-0.05$.  I choose an initial CE surface with an
   expansion of $\lambda=0.07$.  In each snapshot of the series I
   reduce $\lambda$ in steps of $0.01$ until the corresponding surface
   cannot be located any more.  In this case, the solver quickly
   aborts with an error code, which takes no more time than if a
   solution was found.  I then retain the surface from the last
   successful solution.

\begin{figure}
\begin{center}
\includegraphics{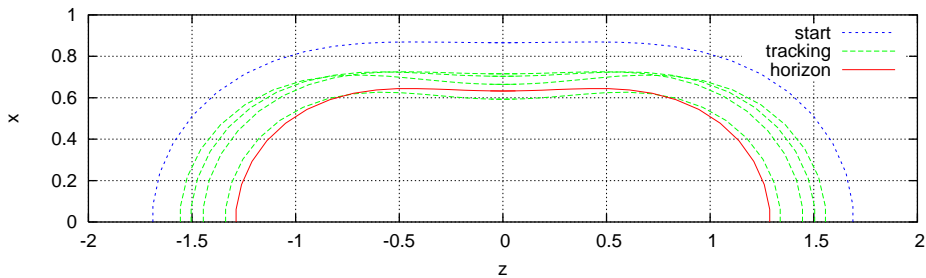}
\end{center}
\caption{
   CE surfaces with decreasing expansions $\lambda$ are tracked during
   a pseudo-evolution of Brill--Lindquist data until an apparent
   horizon is found.  Each surface lives in a different spacelike
   hypersurface corresponding to a different ``time'', where the black
   holes are in different locations.  (Note the directions of the
   coordinate axes.)}
\label{fig:bl2m}
\end{figure}

   Figure~\ref{fig:bl2m} shows the result of this example run, and
   table~\ref{tab:bl2m} lists the black hole locations $z$ and the
   associated minimum expansions $\lambda$.  At $z=0.75$, $\lambda$
   can be reduced to $0$, so that a common apparent horizon with a
   mass of $M \approx 1.976$ is detected.  This value of $z$ is indeed
   close to where the common apparent horizons forms; see e.g.\
   \cite{alcubierre2000}.  The last CE surface in the series that is
   not yet an apparent horizon (at $z=0.80$) has $\lambda=0.01$.  This
   surface is already very close in shape to the apparent horizon.
   Applying the expression for the irreducible mass $M=R/2$ to this
   last non-horizon surface yields a guess of $M \approx 1.968$, which
   differs by less than $1\%$ from the horizon mass.  This indicates
   that pre-tracking does not only speed up locating common horizons,
   but can also be used to extract approximate physical information.

\begin{table}
\caption{
   Parameters of the tracked CE surfaces shown in
   figure~\ref{fig:bl2m}.  $\pm z$ are the coordinate locations of the
   two singularities, $\lambda$ is the expansion of the CE surface,
   $R$ its areal radius.  For $\lambda=0$, $R/2$ is the irreducible
   mass.}
\label{tab:bl2m}
\begin{indented}
\item[]
\begin{tabular}{@{}llll}\br
$z$    & $\lambda$ & $R$       & $R/2$     \\\mr
$1.00$ & $0.07$    & $3.97820$ & $1.98910$ \\
$0.95$ & $0.05$    & $3.92198$ & $1.96099$ \\
$0.90$ & $0.04$    & $3.93548$ & $1.96774$ \\
$0.85$ & $0.03$    & $3.94254$ & $1.97127$ \\
$0.80$ & $0.01$    & $3.93648$ & $1.96824$ \\
$0.75$ & $0.00$    & $3.95218$ & $1.97609$ \\\br
\end{tabular}
\end{indented}
\end{table}

\section{Summary}
\label{sec:summary}

   Apparent horizons and other CE surfaces provide valuable insight
   into spacetimes that are given numerically.  Apparent horizons
   provide mass and spin estimates for black holes; they are
   indicators for event horizons, and help keep track of the location
   of singularities.  CE surfaces are a generalisation of apparent
   horizons.  Empirically, they lead to a foliation of a spacelike
   hypersurface outside a small interior region.  They can be used to
   compare spacetimes that are given in different coordinate systems
   (but with the same slicing), and provide a measure for the
   coordinate distortions of the three-metric.

   The presented CE surface finding algorithm provides an efficient
   and robust way to find apparent horizons and other CE surfaces, and
   also a fast way to track CE surfaces during a numerical evolution,
   helping to find new apparent horizons as they appear.

\ack

   I wish to thank to my collaborators Mijan Huq, Badri Krishnan,
   Pablo Laguna, and Deirdre Shoemaker for countless inspiring
   discussions.  Jan Metzger provided much information regarding CE
   surfaces, minimisation procedures, and curvature flow.  Ian Hawke
   and Scott Hawley gave helpful remarks regarding gauge conditions.
   While working at this project, I was financed by the SFB 382 \glqq
   Verfahren und Algorithmen zur Simulation physikalischer Prozesse
   auf H\"ochstleistungsrechnern\grqq\ of the DFG.  Several visits to
   Penn State University were supported by the NSF grants PHY-9800973
   and PHY-0114375.

\section*{References}

\bibliographystyle{unsrt}
\bibliography{ah}

\end{document}